\documentclass[pra,aps,showpacs,nofootinbib,preprint]{revtex4}

\begin{document}

\preprint{FERMILAB-PUB-05-242-A}

\title{On cosmic acceleration without dark energy}  

\author{Edward W. Kolb}\email{rocky@fnal.gov}
\affiliation{Particle Astrophysics Center, Fermi
       	National Accelerator Laboratory, Batavia, Illinois \ 60510-0500, USA \\
       	and Department of Astronomy and Astrophysics, Enrico Fermi Institute,
       	University of Chicago, Chicago, Illinois \ 60637-1433 USA}
\author{Sabino Matarrese}\email{sabino.matarrese@pd.infn.it}
\affiliation{Dipartimento di Fisica ``G.\ Galilei'' Universit\`{a} di Padova, 
        INFN Sezione di Padova, via Marzolo 8, Padova I-35131, Italy}
\author{Antonio Riotto}\email{antonio.riotto@pd.infn.it}
\affiliation{INFN Sezione di Padova, via Marzolo 8, I-35131, Italy}


\begin{abstract}
We elaborate on the proposal that the observed acceleration of the
Universe is the result of the backreaction of cosmological
perturbations, rather than the effect of a negative-pressure
dark-energy fluid or a modification of general relativity.  Through
the effective Friedmann equations describing an inhomogeneous Universe
after smoothing, we demonstrate that acceleration in our local Hubble
patch is possible even if fluid elements do not individually undergo
accelerated expansion.  This invalidates the no-go theorem that there
can be no acceleration in our local Hubble patch if the Universe only
contains irrotational dust. We then study perturbatively the time
behavior of general-relativistic cosmological perturbations, applying,
where possible, the renormalization group to regularize the dynamics.
We show that an instability occurs in the perturbative expansion
involving sub-Hubble modes. Whether this is an indication 
that acceleration in our Hubble patch originates from the backreaction 
of cosmological perturbations on observable scales requires a fully 
non-perturbative approach.   
\end{abstract} 


\maketitle
\section{Introduction}
Recent observations of the expansion history of the Universe indicate
that the Universe is presently undergoing a phase of accelerated
expansion \cite{original,acceleratedreview}. The accelerated expansion
is usually interpreted as evidence either for a ``dark energy'' (DE)
component to the mass-energy density of the Universe, or for a
modification of gravity at large distances.  In this paper we explore
another possibility, namely that the accelerated expansion is due to
the presence of inhomogeneities in the Universe.

In the homogeneous, isotropic, Friedmann-Robertson-Walker (FRW)
cosmology, the acceleration (or deceleration) of the expansion may be
expressed in terms of a dimensionless parameter $q$, proportional to
the second time derivative of the cosmic scale factor $a$. It is
uniquely determined in terms of the relative densities and the
equations of state of the various fluids by (overdots denote time
derivatives),
\begin{equation}
\label{qdef}
q \equiv -\frac{\ddot{a}a}{\dot{a}^2}
=\frac{1}{2}\Omega_{\rm TOTAL}+\frac{3}{2}\sum_i\,w_i\,\Omega_i ,
\end{equation}
where $\Omega_{\rm TOTAL}$ is the total density parameter and the
factors $\Omega_i$ are the relative contributions of the various
components of the energy density with equation of state
$w_i=P_i/\rho_i$ ($P_i$ and $\rho_i$ being the pressure and energy
density of $i$-th fluid).  The expansion accelerates if $q<0$.
Observations seem to require DE with present values $w_{DE}\sim -1$
and $\Omega_{DE}\sim 0.7$ \cite{rp}. The negative value of $w_{DE}$,
indicating a violation of the energy condition $w>-1/3$
\cite{hawking}, is usually interpreted as the effect of a mysterious
dark energy fluid of unknown nature or a cosmological constant of
surprisingly small magnitude.

The existence of a negative-pressure fluid or a cosmological constant
would have profound implications for physics as well as
cosmology. While the observational evidence for the acceleration of
the Universe is now compelling, it is important to keep in mind that
the evidence for dark energy is indirect; it is {\em inferred} from
the observed time evolution of the expansion rate of the Universe.
What is known is that the expansion history of the Universe is not
described by the expansion history of an Einstein--de Sitter Universe
(a spatially flat, matter-dominated FRW model).  While such a
departure may be caused by dark energy, there are other possibilities.
One possibility is that general relativity is not a good description
of gravity on large distance scales.  Another possibility is that the
Universe is matter-dominated and described by general relativity, and
the departure of the expansion rate from the Einstein--de Sitter model
is the result of back reactions of cosmological perturbations.  This
explanation is the most conservative, since it assumes neither a
cosmological constant, a negative-pressure fluid, nor a modification
of general relativity.

In this paper we explore the possibility that backreactions of
cosmological perturbations is the source of the accelerated expansion
\cite{KMNR,oldKMNR,japan,oldrasanen,rasanen,alessio,sw,bmr,rtolman}. 
The idea is as follows. We know there exist cosmological
perturbations; after all, the Universe is inhomogeneous.  To describe
the time evolution of a patch of the Universe as large as our local
Hubble radius one has to construct the effective dynamics from which
observable average properties can be inferred. Of course, this implies
a scale-dependent description of inhomogeneities.  Suppose further
that our Universe is filled with pressureless matter and no DE.  If
inhomogeneities evolve with time, a local observer would infer that
our Universe is not expanding as a homogeneous and isotropic FRW
matter-dominated Universe with Hubble rate $H(t)\propto t^{2/3}$,
where $t$ is cosmic time. On the contrary, the Universe would appear
to have an expansion rate with a time evolution that depends on the
nature and time evolution of these perturbations. Potentially, this
could lead to an accelerated expansion.

Our idea is actually intimately connected with the general problem of
how the nonlinear dynamics of cosmological perturbations on small
scales may affect the large-scale ``background'' geometry.  Let us
start by discussing this issue in some generality.  

The standard approach to cosmology is based both on observational
facts, such as the near-perfect isotropy of the Cosmic Microwave
Background (CMB) radiation, and on an {\it a priori} philosophical
assumption, the so-called Cosmological or Copernican Principle.
According to the Cosmological Principle all comoving cosmic observers
at a given cosmic time should see identical properties around them
(isotropy around all cosmic observers implies homogeneity, hence the
FRW line element).  The Cosmological Principle allows one to
circumvent our inability to obtain information about the Universe
outside our past light-cone by assuming that a symmetry principle is
valid {\it everywhere}.  By using the Cosmological Principle, we
assume that we are able to determine conditions many Hubble radii away
from us by using observational data within our past light-cone, whose
region of influence is, by definition, limited to one Hubble radius
\cite{ellis84}.

An alternative procedure, dubbed {\it Observational Cosmology}, has
also been proposed. It aims at constructing a cosmological model
solely in terms of observational facts, thereby avoiding any {\it a
priori} assumptions of global symmetry. It dates back to the works by
Kristian and Sachs in 1966 \cite{ks} and Ellis in 1984
\cite{ellis84}. A remarkable feature of this approach is that, by using
Einstein's equations, the dataset observable within our past light-cone
is precisely sufficient to determine the space-time and its matter
content within the same light-cone (see Ref.\ \cite{ellis84} and
references therein).

A crucial ingredient of the Observational Cosmology approach, which is
shared by any realistic cosmological model-fitting procedure, is {\it
smoothing}. Observations tell us that the Universe is far from
homogeneous and isotropic on small scales.  Observationally, we know
that homogeneity, {\it e.g.,} in the galaxy distribution, is only
achieved over some large smoothing scale (see {\it e.g.,} Ref.\
\cite{lahav}). When we refer to homogeneity and isotropy of the
Universe we tacitly assume that spatial smoothing over some suitably
large filtering scale has been applied so that fine-grained details
can be ignored (see in this respect the discussion in Refs.\
\cite{carf,carf2}).  In other words, by the mere assumption that the
same background model can be used to describe the properties of nearby
and very distant objects in the Universe, the smoothing process is
implicit in the way we fit a FRW model to observations.  Cosmological
parameters like the Hubble expansion rate or the energy density of the
various cosmic components are to be considered as volume averaged
quantities: only these can be compared with observations
\cite{ellisnew}.

There is, however, a technical difficulty inherent in any smoothing
procedure. While matter smoothing is somewhat straightforward ({\it
e.g.,} in the fluid description), smoothing of the space-time metric
is more complex and immediately leads to an important and unexpected
feature, pointed out by Ellis \cite{ellis84}.  Let us assume that
Einstein's equations hold on some suitably small scale where the
Universe is highly inhomogeneous and anisotropic. Next, suppose we
smooth over some larger scale.  Einstein's equations are nonlinear:
smoothing and evolution ({\it i.e.,} going to the field equations)
will not commute.  Hence, the Einstein tensor computed from the
smoothed metric would generally differ from that computed from the
smoothed stress-energy tensor. The difference is a tensor appearing on
the right-hand side (RHS) of Einstein's equations that leads to an
extra term in the effective Friedmann equations describing the
dynamics of the smoothing domain.

How can this fact be related to the acceleration of the Universe? The
answer is that this extra source term need not satisfy the usual
energy conditions (according to which our Universe can only
decelerate) even if the original matter stress-energy tensor does. The
fact that the effective stress-energy tensor emerging after smoothing
could lead to a violation of the energy conditions was originally
recognized by Ellis \cite{ellis84}.\footnote{A closely related
discussion can also be found in Ref.\ \cite{turb}.}  As we will
discuss in Sec.\ \ref{effectivefe}, explicit calculations of the
effective Friedmann equations \cite{buchert1,buchert2,japan} confirm
that acceleration is indeed possible even if our Universe is filled
solely with matter.

A closely related question is what is the appropriate scale for which
the smoothing procedure can fit the standard picture of a homogeneous
and isotropic Universe on large scales?  Our choice will be that of
smoothing over a volume of size comparable with present-day Hubble
volume. The precise size of the averaging volume does not matter,
provided it is large enough that the fair sample hypothesis applies,
{\it i.e.,} that volume averages yield an accurate approximation of
statistical ensemble averages. We will nonetheless refer to
scales within (outside) the averaging domain as ``sub-Hubble''
(``super-Hubble'') perturbations.  In doing this we are however
promoting our super-Hubble, or ``zero''-mode, to the role of FRW-like
background.

The next question is what are the scales that determine the dynamics
of our local background. To answer this question we have to recall
what happens in the standard FRW models. The evolution of the global
scale factor $a(t)$, the zero-mode of FRW models, is fully determined
by the matter content of the Universe through the value of
$\Omega_{\rm TOTAL}$ and $\Omega_i$, and via the equation of state
$w_i$ of its components.  That is where microphysics enters the
game. In other words, in the standard FRW picture the evolution of the
Universe as a whole is determined by the dynamics of matter on
sub-Hubble scales.  Similarly, in the Observational Cosmology approach
the dynamics of our local background must be determined by the
observed behavior of matter inhomogeneities within our
past-light-cone. That is where the backreaction of sub-Hubble
inhomogeneities enters the game.  This picture will become clear in
Sec.\ \ref{effectivefe}, where we will introduce two scalars, the
so-called kinematical backreaction $Q_D$ and the mean spatial
curvature $\langle R\rangle_D$, that enter the expression for the
energy density and pressure in the effective Friedmann equations
governing the mean evolution of our local domain $D$.  The crucial
point is that in the fully general relativistic framework these two
scalars are linked together by an integrability condition (which has
no analogue in the Newtonian context), whose solution provides the
effective equation of state of the backreaction. In order to solve
this equation and establish the typical size of these terms one needs
a non-perturbative and non-Newtonian approach to the evolution of
cosmological irregularities, as pointed out in
Refs. \cite{ehlers,bks}.

The fact that the average dynamics naturally leads to new terms in the
source implies that it is legitimate to use an effective Friedmann
description, provided one takes into account that the effective
sources of these equations contain the back-reaction terms.

What will result from our analysis is that the evolution of sub-Hubble
perturbations leads to an instability of the perturbative expansion
due to the presence of large contributions which depend on a combination of
Newtonian and post-Newtonian terms. This instability indicates that
the effective scale factor describing the dynamics of our local Hubble
patch is fed by the evolution of inhomogeneities within the Hubble
radius.  This cross-talk between the small-scale dynamics and the
effective average dynamics described by super-Hubble, or
``zero''-mode, playing the role of FRW-like background might be the
crucial ingredient of backreaction that can lead to cosmic
acceleration without dark energy.  In Ref.\ \cite{KMNR} a deviation
from the pure matter-dominated evolution was obtained by a combination
of sub- and super-Hubble modes generated by inflation, the latter
being improperly used to amplify the backreaction. In this paper, we
will show that the deviation from a matter-dominated background is
entirely due to the nonlinear evolution of sub-Hubble modes which may
cause a large backreaction (technically due to the disappearance of
the filter modeling the volume average), while the super-Hubble modes
play no dynamical role.

At this point it may be useful to contrast the differences between our approach
dealing with inhomogeneities with the traditional approach.  In the traditional
approach one averages over inhomogeneities and forms a time-dependent average
energy density $\langle\rho(\vec{x},t)\rangle$ (although the standard procedure
is to calculate averages with the unperturbed spatial metric!).  One then uses
for the dynamics of the ``zero-mode'' [$a(t)$] the dynamics of a homogeneous
universe with energy density $\rho(t) = \langle\rho(\vec{x},t)\rangle$.  One
then regards inhomogeneities as a purely ``local'' effect, for instance,
leading to peculiar velocities.  In this approach inhomogeneities can not
result in acceleration.

In our approach, we take note of the fact that the expansion rate of an
inhomogeneous universe of average density $\langle\rho(\vec{x},t)\rangle$
(using the inhomogeneous spatial metric to calculate the spatial average!)
is not the same as the expansion rate of a universe with the same average
density. In order to account for this we encode the expansion dynamics
into a new zero mode (or scale factor) $a_D(t)$ (which will be properly
defined in the next section).  It is the dynamics of this renormalized
scale factor that is best used to calculate observables and will determine
whether the Universe accelerates.  In our approach the effect of
short-wavelength inhomogeneities is not just a local effect, but
renormalizes the long-wavelength dynamics.

Our paper is organized as follows. In Sec.\ \ref{effectivefe} we
summarize the  effective Friedmann  description of an
inhomogeneous Universe after smoothing.  In Sec.\ \ref{appearins} we
discuss how acceleration in our Hubble patch can result from the
backreaction of perturbations.  Conclusions are drawn in Sec.\
\ref{conclusions}. The Appendix presents the main results of a
fourth-order gradient-expansion technique.

\section{Effective Friedmann equations in an inhomogeneous Universe
\label{effectivefe}}
The goal of this section is to compute the time dependence of the
local expansion rate of the Universe. For a generic fluid we may take
the four-velocity to be $u^\mu=(1,\vec{0})$, which amounts to saying
that a local observer is comoving with the energy flow of the
fluid. For the case of irrotational dust considered in this paper we
have the freedom to work in the synchronous and comoving gauge with
line element
\begin{equation}
\label{oo}
ds^2=-dt^2 + h_{ij}({\bf x},t) dx^i dx^j  ,
\end{equation}
where $t$ is cosmic time.  

A fundamental quantity in our analysis is the velocity gradient
tensor, which is purely spatial and symmetric because of
irrotationality.  It is defined as
\begin{equation}
\Theta^i_{\ j}=u^i_{\ ;j} = \frac{1}{2} h^{ik} \dot{h}_{kj}.
\end{equation}
Here dots denote derivatives with respect to cosmic time.  The tensor
$\Theta^i_{\ j}$, represents the {\it extrinsic curvature} of the
spatial hypersurfaces orthogonal to the fluid flow.  It may be written
as
\begin{equation}
\Theta^i_{\ j} = \Theta \, \delta^i_{\ j} + \sigma^i_{\ j} .
\end{equation}
Here $\Theta$ is called the {\it volume-expansion scalar}, reducing to
$3H$ ($H$ is the usual Hubble rate) in the homogeneous and isotropic
FRW case.  The traceless tensor $\sigma^i_{\ j}$ is called the {\it
shear}.

The evolution equations for the expansion and the shear come from the
space-space components of Einstein's equations (see {\it e.g.,} Ref.\
\cite{mater}). They read, respectively, ($\rho$ is the mass density, 
$R$ and $R^i_{~j}$ are the spatial Ricci scalar and tensor respectively of
comoving space-like hypersurfaces)
\begin{eqnarray}
\label{thetaevol}
\dot{\Theta} + \Theta^2 + R & = & 12 \pi G\, \rho , \\
\label{shearevol}
\dot{\sigma}^{i}_{\ j} + \Theta \sigma^i_{\ j} + 
R^i_{\ j} - \frac{1}{3} R \delta^i_{\ j} & = & 0 . 
\end{eqnarray}
The $0-0$ component of Einstein's equations is the {\it energy
constraint}
\begin{equation} 
\label{ec}
\frac{2}{3}\Theta^2 - 2\sigma^2 + R = 16 \pi G\, \rho , 
\end{equation}
where $\sigma^2 \equiv \frac{1}{2}\sigma^i_{\ j} \sigma^j_{\ i}$. 
The $0-i$ components yield the {\it momentum constraint} 
\begin{equation}
\label{mc}
\sigma^i_{\ j|i} - \frac{2}{3}\Theta_{,j} = 0 ,
\end{equation}
where the vertical bar denotes covariant differentiation in the
three-space with metric $h_{ij}$.  The mass density, in turn, can be
obtained from the continuity equation
\begin{equation}
\label{dceq}
\dot\rho = - \Theta \rho  ,
\end{equation}
whose solution reads
\begin{equation}
\label{dc}
\rho = \rho_0 \left(h/h_0\right)^{-1/2} ,
\end{equation}
where $h\equiv {\rm det}\, h_{ij}$ and the initial conditions have
been arbitrarily set at the present time $t_0$.  Combining the
expansion evolution equation with the energy constraint gives the {\it
Raychaudhuri equation},
\begin{equation}
\dot{\Theta} + \frac{1}{3} \Theta^2 + 2 \sigma^2 + 
4\pi G\, \rho =0  . 
\end{equation}
From the latter equation it is straightforward to verify that
irrotational pressure fluid elements cannot locally undergo
accelerated expansion.  (This point was emphasized by Hirata and
Seljak \cite{seljak}.) Indeed, defining a local deceleration parameter
and using the Raychaudhuri equation, one finds
\begin{equation}
q \equiv - \left( 3 \dot{\Theta} + \Theta^2\right)/\Theta^2=
6 (\sigma^2 + 2 \pi G\rho)/\Theta^2 \geq 0 .
\end{equation}

While it is true that locally the expansion does not accelerate, it is
incorrect to assume that acceleration can not occur when the fluid is
coarse-grained over a finite domain. The reason is trivial: the time
derivative of the average of $\Theta$ and the average of the time
derivative of $\Theta$ are not the same because of the time dependence
of the coarse-graining volume.

Let us denote the coarse-grained value of a quantity ${\cal F}$ by its
average over a spatial domain $D$:\footnote{Notice that one is not
allowed to define the mean cosmological parameters only through an
average over directions \cite{flanagan,seljak} as cosmological
observables, such as the Hubble rate, receive unacceptably large
corrections from the same Newtonian terms which become harmless
surface terms when averaging over a large volume
\cite{oldKMNR}. We acknowledge discussions with U.\ Seljak about this issue.} 
\begin{equation}
\label{coarsegrain}
\langle {\cal F} \rangle_D = \frac{\int_D  
\sqrt{h} \,{\cal F}\, d^3\!x} {\int_D 
\sqrt{h}\,d^3\!x}  .
\end{equation} 
We will take the domain to be comparable with the size of the present
Hubble volume\footnote{The correct definition of our comoving Hubble
radius is $R_H(t_0) = e^{-\Psi_{\ell0}} \int^{t_0} dt
e^{\Psi_\ell(t)}$.}.

A first important property follows directly from the smoothing
procedure itself: for a generic function ${\cal F}$ one has
\cite{buchert1,buchert2}
\begin{equation}
\langle {\cal F} \rangle_D^\cdot - \langle \dot{{\cal F}} \rangle_D =
\langle {\cal F}\Theta \rangle_D - \langle \Theta\rangle_D\langle{\cal F}
\rangle_D ,
\end{equation}
where we have not considered terms originating from the peculiar motion of 
the boundary, since we will eventually consider only comoving domains in what  
follows. 
In particular, for the local expansion rate one finds
\begin{equation}
\langle \Theta \rangle_D^\cdot = \langle \dot{\Theta} \rangle_D +
\langle \Theta^2 \rangle_D - \langle \Theta\rangle_D^2\geq \langle 
\dot{\Theta}\rangle_D \,. 
\end{equation}
Although $\langle \dot{\Theta} \rangle_D 
\leq -\frac{1}{3}\langle \Theta^2\rangle_D \leq 0$, 
the coarse-grained deceleration parameter $q_D\equiv 
-3\langle\Theta\rangle_D^\cdot/\langle\Theta\rangle_D^2-1$
is related to 
$\langle\Theta\rangle_D^\cdot$,
which is not the same as $\langle \dot{\Theta} \rangle_D$.  It is
precisely this commutation rule that allows for the possibility of
acceleration in our local patch in spite of the fact that fluid
elements cannot individually undergo accelerated expansion.  This
simple argument circumvents the {\it no-go theorem} adopted in  Refs.\
\cite{flanagan,seljak} (and later  in \cite{giov}), according to 
which there can be no acceleration in our local Hubble patch if the
Universe only contains irrotational dust.

Indeed, let us follow the work of Buchert \cite{buchert1,buchert2} and
define a dimensionless scale factor
\begin{equation}
a_D(t) \equiv \left(\frac{V_D}{V_{D_0}}\right)^{1/3} ; \qquad 
V_D   = \int_D\, \sqrt{h}\,d^3\!x ,
\end{equation}
where $V_D$ is the volume of our coarse-graining domain (the subscript
``$0$'' denotes the present time).  As an averaging volume we may take
a large comoving domain, so that $D$ is constant in time.  Alternative
choices are however possible (see, {\it e.g.,} Ref.\ \cite{elsto} for
a discussion of different averaging procedures).\footnote{One could
alternatively average over a volume of size of the order of the
instantaneous particle horizon. The effective dynamics in this case
will be accompanied by a stochastic source originated by the
statistical nature of the perturbations.}

The coarse-grained Hubble rate $H_D$ will be 
\begin{equation}
H_D = \frac{\dot{a}_D}{a_D}=\frac{1}{3}\langle\Theta\rangle_D .
\end{equation}
(Notice that with such a coarse-graining, $H_D$ coincides with the
quantity $\overline{H}$ defined in Ref.\ \cite{KMNR}).  By properly
smoothing Einstein's equations over the volume $V_D$, one obtains
\cite{buchert1,buchert2}
\begin{eqnarray}
\label{FR1}
\frac{\ddot{a}_D}{a_D}& = & -\frac{4\pi G}{3}
\left(\rho_{\rm eff}+ 3 P_{\rm eff}\right) , \\
\label{FR2}
\left(\frac{\dot{a}_D}{a_D}\right)^2 & = & \frac{8\pi G}{3}\rho_{\rm eff} ,
\end{eqnarray}
where we have defined effective energy density and pressure terms 
\begin{eqnarray}
\label{effenergy}
\rho_{\rm eff} &=& \langle \rho\rangle_D-\frac{Q_D}{16\pi G}-
\frac{\langle R\rangle_D}{16\pi G}  \\
\label{effpressure}
P_{\rm eff}&=&-\frac{Q_D}{16\pi G}+
\frac{\langle R\rangle_D}{48\pi G}  ,
\end{eqnarray}
and we have introduced the {\it kinematical backreaction} 
\begin{equation}
\label{list}
Q_D=\frac{2}{3}\left(\langle\Theta^2\rangle_D-\langle\Theta\rangle_D^2\right)
-2\langle\sigma^2\rangle_D . 
\end{equation}

From the effective Friedmann equations, Eqs.\ (\ref{FR1}) and
(\ref{FR2}), obtained by Buchert in Ref.\ \cite{buchert1}, one
immediately obtains the continuity equation for our effective fluid
\begin{equation}
\label{continuity}
\dot{\rho}_{\rm eff} = -3H_D\left(\rho_{\rm eff}+ P_{\rm eff} \right) . 
\end{equation}
Note that the smoothed continuity equation differs from the local
continuity equation.  On the other hand, owing to the fact that our
coarse-graining volume is comoving with the mass flow, mass
conservation is preserved by the smoothing procedure, implying
\begin{equation} 
\label{mass}
{\langle \rho\rangle}_D^\cdot = - 3 H_D \langle \rho\rangle_D . 
\end{equation}

The two quantities $Q_D$ and $\langle R\rangle_D$ are not independent.
This can be seen by taking the derivative of Eq.\ (\ref{FR2}) and
using Eq.\ (\ref{mass}).  The consistency of the system of Eqs.\
(\ref{FR1}), (\ref{FR2}), and (\ref{mass}) requires that $Q_D$ and
$\langle R\rangle_D$ satisfy the {\it integrability condition}
\cite{buchert1}
\begin{equation}
\label{relation}
\left(a_D^6Q_D\right)^\cdot+a_D^4\left(a_D^2\langle R\rangle_D\right)^\cdot=0 .
\end{equation}

One should stress that the latter equation, {\it i.e.,} the link
between kinematical backreaction $Q_D$ and mean curvature $\langle
R\rangle_D$, is a genuine General Relativistic (GR) effect, having no
analogue in Newtonian theory, as the curvature $R$ of comoving
hypersurfaces vanishes identically in the Newtonian limit
\cite{ellis,mater,buchert1}, implying that there exist globally flat Eulerian
coordinates $X^i$.  Indeed, in the Newtonian case, it is immediate to
verify that $Q_D$ is {\it exactly} (i.e., at any order in perturbation
theory) given by the volume integral of a total-derivative term in
Eulerian coordinates \cite{ehlers},
\begin{equation}
Q^{\rm Newtonian}_D = \left\langle \nabla \cdot \left[{\bf u} 
\left(\nabla \cdot 
{\bf u}\right) - \left({\bf u} \cdot \nabla \right) {\bf u} \right] 
\right\rangle_D,
\end{equation} 
where $\bf u$ is the peculiar velocity, so that by the Gauss theorem
it can be transformed into a pure boundary term.  It is precisely by
this reason that any analysis of backreaction based on the Newtonian
approximation, such as that recently performed in v1 of Ref.\ \cite{fry}, is
not relevant: it will invariably lead to a tiny effect,
and to the absence of any acceleration.  Indeed, if inhomogeneities
only exist on scales much smaller than our Hubble radius and if
peculiar velocities are small on the boundary of our Hubble patch
then, {\it within the Newtonian approximation}, the standard FRW
matter-dominated model can be applied without any substantial
correction from the backreaction \cite{ehlers}. Such a drawback of the
Newtonian approximation was also noticed in Refs.\
\cite{oldrasanen,oldKMNR}.
This exact result demonstrates that in order to deal with the
backreaction, going beyond the Newtonian approximation is mandatory,
as also stressed in Ref. \cite{japan}. Studies of the average
dynamics including the lowest post-Newtonian gradient terms in the 
weak field limit were considered in Refs. \cite{newton} and v2 of 
\cite{fry}. However,  further and 
more sizeable terms are expected to   contribute to the backreaction
once the effective dynamics of the system (including the kinematical
backreaction) is considered.
We will come back to this issue in  subsection IIIC.

The GR integrability condition makes it clear how acceleration in our
local Hubble patch is possible. Indeed, it is immediate to realize
that the general condition for acceleration in a domain with mean
density $\langle \rho\rangle_D$ is
\begin{equation}
Q_D > 4\pi G\, \langle\rho\rangle_D .
\end{equation} 
Moreover, a particular solution of the integrability condition for
constant $Q_D$ and $\langle R\rangle_D$ is
\begin{equation}
Q_D=-\frac{1}{3}\langle R\rangle_D= \textrm{const.} ,
\end{equation}  
which, for negative mean curvature mimics a cosmological constant,
$\Lambda_{\rm eff}=Q_D$. More in general, if $Q_D$ is positive, it may
mimic a dynamical dark energy or quintessence. 

So far the considerations have been rather general. Now we write the
spatial metric in the general form
\begin{equation} 
h_{ij} \equiv a^2(t)e^{-2 \Psi({\bf x},t)} \left[\delta_{ij} + 
\chi_{ij}({\bf x},t)\right] ,
\end{equation} 
where $a(t) \propto t^{2/3}$ is the usual FRW scale-factor for a flat,
matter-dominated Universe and the traceless tensor $\chi_{ij}$
contains the remaining modes of the metric, namely one more scalar, as
well as vector and tensor modes.\footnote{Indices of $\chi_{ij}$ will
be raised by the Kronecker symbol: $\chi^i_{\ j} \equiv
\delta^{ik}\chi_{kj}$ and $\chi^{ij} \equiv
\delta^{ik} \delta^{jl} \chi_{kl}$.} Next, when we consider the expansion in
some domain $D$, we can split the gravitational potential $\Psi$ into
two parts: $\Psi = \Psi_\ell + \Psi_s$, where $\Psi_\ell$ is the
long-wavelength mode and $\Psi_s$ is a collection of short-wavelength
modes.  Of course ``long'' and ``short'' describe wavelengths compared
to the size of the domain $D$.  We can easily take into account the
effect of $\Psi_\ell$ by noting that within $D$ the factor
$e^{-2\Psi_\ell}$ is a space-independent conformal rescaling of the
spatial metric. Let us then write
\begin{equation}
\label{conformal} 
h_{ij} = a^2(t)e^{-2\Psi_\ell(t)} \widetilde{h}_{ij}({\bf x},t) ,  
\end{equation} 
with $\widetilde{h}_{ij} = e^{-2\Psi_s({\bf x},t)}\left[\delta_{ij} +
\chi_{ij}({\bf x},t)\right]$.  The expansion scalar and shear then become 
\begin{eqnarray} 
\Theta & = & 3\frac{\dot{a}}{a} + \widetilde{\Theta} - 3 \dot{\Psi}_\ell ,  
\nonumber \\ 
\sigma^i_{\ j} & = & \widetilde{\sigma}^i_{\ j} ,  
\end{eqnarray} 
where $\widetilde{\Theta}$ and $\widetilde{\sigma}^i_{\ j}$ are
calculated with $\widetilde{h}_{ij}$. Note that $\widetilde{\Theta}$
and $\widetilde{\sigma}^i_{\ j}$ do not depend {\em explicitly} on
$\Psi_\ell$.  It should be kept in mind that the local expansion rate
is $\Theta$, not $\widetilde{\Theta}$.

The Ricci tensor of comoving space-like hypersurfaces is given by
\begin{equation}
R^i_{\ j}  = a^{-2}e^{2\Psi_\ell} \left[\widetilde{R}^i_{\ j} 
+ \widetilde{\nabla}^i \widetilde{\nabla}_j \Psi_\ell 
+ \widetilde{\nabla}^2 \Psi_\ell \delta^i_{\ j} 
+ \widetilde{\nabla}^i \Psi_\ell \widetilde{\nabla}_j \Psi_\ell 
- \widetilde{\nabla}^k \Psi_\ell \widetilde{\nabla}_k \Psi_\ell \delta^i_{\ j} 
\right] , 
\end{equation}
where $\widetilde{R}^i_{\ j}$ is the Ricci scalar of the metric
$\widetilde{h}_{ij}$ and the symbol $\widetilde{\nabla}_i$
denotes the covariant derivative in the 3-space with metric
$\widetilde{h}_{ij}$. For the Ricci scalar we find
\begin{eqnarray}
R & = & a^{-2}e^{2\Psi_\ell} \left[\widetilde{R} + 
4 \widetilde{\nabla}^2 \Psi_\ell  
- 2 \widetilde{\nabla}^k \Psi_\ell \widetilde{\nabla}_k 
\Psi_\ell \right] \\
\label{r}
\langle R\rangle_D & = & a^{-2}e^{2\Psi_\ell}  
\left\langle\widetilde{R} + 4\widetilde{\nabla}^2\Psi_\ell - 
2\widetilde{\nabla}^k \Psi_\ell \widetilde{\nabla}_k \Psi_\ell 
\right\rangle_D .
\end{eqnarray}

Notice that $a_D$ coincides with the scale factor
$\overline a$ adopted in Ref.\ \cite{KMNR}, provided we take
\begin{equation}
\label{defabar}
a_D(t) =a(t)\, e^{-\Psi_\ell(t) + \Psi_{\ell 0}}
\end{equation}
with 
\begin{equation}
\label{psilong}
\Psi_\ell({\bf x}_{\rm obs},t) \equiv \ln\, a- \frac{1}{3} \ln 
\left(\int_D\, \sqrt{h}\,d^3\!x \right) + {\rm const.} ,
\end{equation}
where the residual dependence of $\Psi_\ell$ on the spatial coordinate
${\bf x}_{\rm obs}$ labels the individual comoving volume patch, {\it i.e.,}
the specific {\it cosmic observer} we are considering. Using Eq.\
(\ref{dc}) we can rewrite Eq.\ (\ref{psilong}) in the form
\begin{equation}
\label{psilong1}
\Psi_\ell(t) = 
- \frac{1}{3} \ln 
\left\langle \left(1+ \delta_{\rm FRW}\right)^{-1} 
\right\rangle_{D_{\rm in}} + {\rm const.}  ,
\end{equation}
where $\delta_{\rm FRW}$ is the {\it density contrast} with respect to
the mean density of a flat, matter-dominated FRW (Einstein-de Sitter)
model, defined through $\rho=\left(1+\delta_{\rm FRW}\right)/
\left(6\pi G\,t^2 \right)$, and ``in'' denotes the initial time, which
for simplicity we have taken to coincide with the end of
inflation. For any quantity ${\cal F}$
\begin{equation}
\langle {\cal F} \rangle_{D_{\rm in}} = 
\frac{\int_D\sqrt{ h_{\rm in}}\, {\cal F}  \, d^3\!x}
{\int_D\sqrt{h_{\rm in}} \, d^3\!x} .
\end{equation}
By inspecting Eq.\ (\ref{psilong1}) one immediately realizes that
acceleration may be achieved in those Hubble patches where the mean
{\it rarefaction factor} $\langle \left(1+\delta_{\rm FRW}\right)^{-1}
\rangle_{D_{\rm in}}$ grows fast enough to compensate for the
Einstein-de Sitter expansion rate $H=2/3t$. Note also that
the integral defining $\Psi_\ell$ is dominated by the dynamics of the
most underdense fluid elements, not by the densest ones, so the
complex dynamics of highly nonlinear mass concentrations never enters
the calculation; by the same reasoning, any intrinsic limitation
related to caustic formation would not affect the validity of our
backreaction treatment.

The kinematical backreaction $Q_D$ is non-vanishing and gets 
contributions only from $\widetilde{\Theta}$ and from the shear 
$\widetilde{\sigma}^i_{\ j}$:
\begin{equation}
\label{q}
Q_D=\frac{2}{3}\langle\widetilde{\Theta}^2\rangle_D 
- 2\langle\widetilde{\sigma}^2\rangle_D, 
\end{equation}
where we used the fact that $\langle\widetilde{\Theta}\rangle_D =0$ by
construction.  

In order to have a qualitative understanding of why acceleration can
be the natural outcome of the backreaction, let us rewrite the mean
expansion rate in terms of the peculiar expansion rate $\theta$, defined  
by $\Theta=3H+\theta$:
\begin{equation}
H_D = \frac{2}{3t} + \frac{1}{3} 
\frac{\langle\left(1+\delta_{\rm FRW}\right)^{-1} \theta 
\rangle_{D_{\rm in}}} 
{\langle\left(1+\delta_{\rm FRW}\right)^{-1}\rangle_{D_{\rm in}}} , 
\end{equation}
which shows that the {\it mean} expansion rate receives a correction with
respect to the FRW background value by the peculiar expansion rate of
mostly underdense regions (where $\theta$ is largest), which 
give the largest contribution to the average. However, as we already noticed,
acceleration may be achieved when $Q_D$ is positive and large
enough. This requires a large variance of the volume expansion rate
within the averaging domain. At early times, when perturbations are
linear, $\Theta$ is narrowly peaked around its mean background value
$3H$. When non-linearities set in, the variance increases because of
the {\it simultaneous} presence of largely under- and over-dense
regions (in fact, counting only under-dense regions would reduce  the
variance leading to an under-estimate of  $Q_D$) \cite{comment}.

\section{The appearance of instabilities \label{appearins}}

In order to discuss the dynamics of our local Hubble patch, one may
proceed along two complementary directions. Either one tries to encode
the effect of perturbations into the local scale factor $a_D$ as done
in Ref. \cite{KMNR}, or one may try to construct an effective equation
of state by computing the backreaction terms $Q_D$ and $\langle
R\rangle_D$. We will try to follow a combination of them to see under
which circumstances acceleration in our Hubble patch can be achieved.

\subsection{The effect of super-Hubble modes}
 
The mean curvature generally depends on both sub- and super-Hubble
modes.  Since $Q_D$ does not depend explicitly on perturbations with
wavelengths larger than the Hubble radius, one immediately concludes
that if one considers super-Hubble modes only, $Q_D$ vanishes and from
the integrability condition $\langle R\rangle_D$ scales like
$a_D^{-2}$.  Therefore the effect of pure super-Hubble perturbations
is limited to generating a true local curvature term which may be
important but can not accelerate the expansion of the Universe.

As we already anticipated in the Introduction, the effective scale
factor describing the dynamics of our local Hubble patch is fed by the
evolution of inhomogeneities within the Hubble radius. This cross-talk
between the small-scale dynamics and the effective average dynamics is
a crucial ingredient, as also pointed out in Ref. \cite{KMNR}. Before
coming to this crucial point, let us pause for a moment and show that
there is another more technical way to achieve the same conclusion
about the role played by super-Hubble modes starting from the
spatial-gradient expansion of Einstein equations.

The spatial-gradient expansion is a nonlinear approximation method
suitable to describe the long-wavelength part of inhomogeneities in
the Universe. This scheme is based on the assumption that observables
like the local curvature can be expanded in powers of gradients of the
perturbations. To account for the effect of super-Hubble modes at late
times we may adopt the so-called renormalization group method applied
to the gradient expansion of Einstein equations \cite{nambu}.  This
will result in the renormalized long-wavelength solution to Einstein
equations, valid also at late times until the long-wavelength
perturbations enter the horizon.

Here we sketch a non-perturbative technique to solve Einstein's
equations in an inhomogeneous Universe.  A more detailed presentation
of the method will be presented elsewhere \cite{pilli}.

Our approach makes use of a gradient-expansion approximation (see
Refs.\ \cite{LK,Tom,Sal,Der,Nam,BS,4grad}). The idea is to describe
the dynamics of irregularities in a Universe which contains
inhomogeneities on scales larger than the Hubble radius.  Working in
the synchronous gauge one expands Einstein's equations starting from a
space-dependent ``seed'' metric.  The lowest order solution
corresponds to the so-called long-wavelength approximation, while
adding higher-order gradients leads to a more accurate solution, which
hopefully converges toward the exact one.

The gradient-expansion technique amounts to keeping a finite number of
spatial derivatives.  This approximation technique is non-perturbative
in the sense that by solving for the metric coefficients $\Psi$ and
$\chi_{ij}$ up to $2n$ spatial gradients one obtains terms of any
order in the conventional perturbative expansion containing up to $2n$
gradients. For the purpose of this paper, working to four derivatives
in $\Psi$ and $\chi_{ij}$ will suffice.  Note that because the scalar
$\Psi$ appears in the argument of an exponential in the way we write
the spatial metric, our gradient-expansion sensibly differs from that
used in Refs.\ \cite{LK,Tom,Sal,Der,Nam,BS,4grad}. Even if $\Psi$ is
obtained up to a finite number of spatial gradients, $h_{ij}$ will
necessarily contain gradient terms of any order.

Since the cosmological perturbations are generated during inflation,
it is natural to set initial conditions for the gravitational
perturbations $\Psi$ and $\chi_{ij}$ at the end of inflation
(effectively coinciding with $t=t_{\rm in}=0$). If so, the spatial
metric in the super-horizon regime is given by
$h_{ij}=a^2e^{-2\zeta}\delta_{ij}$ where $\zeta$ is the curvature
perturbation \cite{en}. It is related to the so-called peculiar
gravitational potential $\varphi(\bf{x})$ defined by $\varphi=3\zeta/5$, 
in a matter-dominated Universe. The comoving curvature perturbation 
is constant on super-horizon scales (up to gradients) when only adiabatic 
modes are present and the decaying mode is disregarded. 
Therefore, the initial conditions at $t=0$ are $\Psi_{\rm in} \equiv
\Psi(t=0)=5\varphi/3$ and $\chi_{ij}(t=0)=0$.  Notice that since cosmological
perturbations generated during single-field models of inflation are
very nearly Gaussian with a nearly flat spectrum \cite{en}, we infer
that $\varphi$ should be regarded as a nearly scale-invariant,
quasi-Gaussian random field.

From this analysis we see that $\Psi$ contains at least a
zero-derivative term: this will be our seed metric perturbation, which
is the necessary ingredient for gravitational instability to develop.
The traceless tensor $\chi_{ij}$ has at least two spatial gradients.
The only exception might come from linear or higher-order tensor modes
appearing at the end of inflation.  Nonetheless, accounting for these
contributions does not quantitatively affect our results.

Let us adopt the gradient-expansion in order to 
obtain $\Psi$ and $\chi_i^j$ including up to four gradients 
of the initial seed potential $\varphi$.
Under this assumption, the inverse spatial
metric can be written as 
\begin{equation}
h^{ij} = a^{-2}e^{2\Psi} \left(\delta^{ij} 
- \chi^{ij} + \chi^i_{\ k}\chi^{kj} \right).
\end{equation} 
Writing $\Theta \equiv 3H + \theta$ where $H=\dot{a}/{a}$, one finds 
(up to higher-derivative terms)
\begin{eqnarray}
\label{truetheta}
\theta & = & -3 \dot{\Psi} - \frac{1}{2} \chi^{kl}\dot{\chi}_{kl} \\
\sigma^i_{\ j} & = & \frac{1}{2} \dot{\chi}^{i}_{\ j} -  
\frac{1}{2} \chi^{ik}\dot{\chi}_{kj}
+ \frac{1}{6} \chi^{lk}\dot{\chi}_{lk} \delta^i_{\ j} . 
\end{eqnarray}

To solve for these quantities one also needs the 3D Ricci tensor and
Ricci scalar of the constant-time hypersurfaces.  We start by
calculating the 3D Christoffel symbols, which read (up to
higher-derivative terms)
\begin{equation}
\Gamma^i_{jk} = - \Psi_{,k} \, \delta^i_{\ j} - \Psi_{,j} \, \delta^i_{\ k}
+  \Psi^{,i} \, \delta_{jk} + \frac{1}{2} 
\left( \chi^i_{\ j,k} + \chi^i_{\ k,j} - \chi^{\ \ ,i}_{jk} \right)
+ \Psi^{,i} \, \chi_{jk} - \Psi_{,l} \, \chi^{il} \delta_{jk}  .  
\end{equation}
The Ricci tensor and Ricci scalar read, respectively, 
\begin{eqnarray}
R^i_{\ j} & \equiv & \frac{{\cal R}^i_{\ j}}{a^2} = 
  \frac{e^{2\Psi}}{a^2}\left[ \Psi^{,i}_{\ ,j} + \nabla^2 \Psi \delta^i_{\ j} 
+ \Psi^{,i} \Psi_{,j} - \left(\nabla\Psi\right)^2\delta^i_{\ j} 
- \chi^{ik}\Psi_{,kj} - \chi^{ik} \Psi^{,k} \Psi_{,j} \right. \nonumber \\ 
& & 
+ \frac{1}{2} \left( \chi^{ik}_{\ ,kj} + \chi^{k\ ,i}_{\ j \  ,k} 
- \nabla^2 \chi^i_{\ j} \right) 
- \Psi^{,kl} \chi_{kl} \delta^i_{\ j} 
- \Psi_{,k} \chi^{kl}_{\ ,l} \, \delta^i_{\ j} \nonumber \\
& & \left. + \frac{1}{2} \Psi^{,k} \left( - \chi^i_{\ k,j} -
\chi^{\ \ ,i}_{kj} + \chi^i_{\ j,k} \right) + 
\Psi_{,k} \Psi_{,l} \chi^{kl} \delta^i_{\ j} \right]  \\
R & \equiv & \frac{\cal R}{a^2} = \frac{e^{2\Psi}}{a^2}
\left[4 \nabla^2 \Psi - 2 
\left(\nabla\Psi\right)^2 + \chi^{ij}_{\ ,ij}
-  4 \chi^{ij}\Psi_{,ij} - 4 \chi^{ij}_{\ ,i}\Psi_{,j} 
+ 2 \chi^{ij}\Psi_{,i} \Psi_{,j} \right]  . 
\end{eqnarray}

The evolution equations for the peculiar volume expansion scalar 
and for the shear  immediately follow from Eqs.\ (\ref{thetaevol}),  
(\ref{shearevol}) and (\ref{ec})
\begin{eqnarray} 
\dot{\theta} + 3 H  \theta + \frac{1}{2} \theta^2 + 
\frac{3}{2}\sigma^2 & = & - \frac{1}{4a^2} {\cal R}  , \nonumber \\
\dot{\sigma}^i_{\ j} + 3 H \sigma^i_{\ j} + \theta \sigma^i_{\ j} & = & - 
\frac{1}{a^2} \left({\cal R}^i_{\ j} - 
\frac{1}{3} {\cal R} \delta^i_{\ j}  \right) .
\end{eqnarray}

Replacing our expressions for $\theta$, $\chi^i_{\ j}$, 
${\cal R}^i_{\ j}$ and ${\cal R}$ in terms of the metric coefficients and 
retaining only terms containing up to  four spatial derivatives, one obtains 
differential equations for $\Psi$ and $\chi_{ij}$, namely 
\begin{eqnarray} 
\label{k}
\ddot{\Psi} + 3H \dot{\Psi} & = &
\frac{3}{2} \dot{\Psi}^2 - \frac{1}{2} H \chi^{kl} 
\dot{\chi}_{kl} - \frac{5}{48} \dot{\chi}^{kl} \dot{\chi}_{kl} - 
\frac{1}{6} \chi^{kl} \ddot{\chi}_{kl} + \frac{1}{12a^2} {\cal R} , \\
\label{kk}
\ddot{\chi}^i_{\ j} + 3H \dot{\chi}^i_{\ j} & = &
3H \left(\chi^{ik} \dot{\chi}_{kj} 
- \frac{1}{3} \chi^{kl} \dot{\chi}_{kl} \delta^i_{\ j} \right)
+ \left(\dot{\chi}^{ik} \dot{\chi}_{kj} 
- \frac{1}{3} \dot{\chi}^{kl} \dot{\chi}_{kl} \delta^i_{\ j} \right)
\nonumber \\
& + & \left(\chi^{ik} \ddot{\chi}_{kj} 
- \frac{1}{3} \chi^{kl} \ddot{\chi}_{kl} \delta^i_{\ j} \right)+
3 \dot{\Psi} \dot{\chi}^i_{\ j} - \frac{2}{a^2} 
\left({\cal R}^i_{\ j} - 
\frac{1}{3} {\cal R} \delta^i_{\ j}  \right)  . 
\end{eqnarray}

These equations can be solved iteratively. With two gradients only,
the conformal Ricci tensor ${\cal R}^i_{\ j}$ and scalar ${\cal R}$
coincide with their initial values, {\it i.e.,} with the curvature of
the seed conformal metric $e^{-2\Psi_{\rm in}}\delta_{ij}$. Up to two
gradients we obtain
\begin{eqnarray}
\label{psi}
\Psi & = & \frac{5}{3}\varphi + \frac{1}{18}\left(\frac{a}{a_0}\right)
\left(\frac{2}{H_0}\right)^2
 e^{10\varphi/3}\left[\nabla^2 \varphi - \frac{5}{6} 
\left(\nabla \varphi\right)^2\right], \\
\label{chi}
\chi^i_{\ j} & = & - \frac{1}{3}\left(\frac{a}{a_0}\right)
\left(\frac{2}{H_0}\right)^2 e^{10\varphi/3}
\left[D^i_{\ j} \varphi + \frac{5}{3} 
\left(\varphi^{,i} \varphi_{,j} - \frac{1}{3} \left(\nabla \varphi\right)^2
\delta^i_{\ j} \right) \right]  ,
\end{eqnarray}
where $D^i_{\ j} \equiv \partial^i\partial_j - \frac{1}{3} 
\nabla^2 \delta^i_{\ j}$. 

From Eq.\ (\ref{psi}) we obtain the volume expansion scalar and the
shear tensor:
\begin{eqnarray}
\label{theta}
\theta & = & - \frac{1}{3}\left(\frac{a}{a_0}\right)^{-1/2}
\left(\frac{2}{H_0}\right) e^{10\varphi/3}  \left[ 
\nabla^2 \varphi - \frac{5}{6} \left(\nabla \varphi\right)^2\right] \\
\label{sigma}
\sigma^i_{\ j} & = & - \frac{1}{3}\left(\frac{a}{a_0}\right)^{-1/2}
\left(\frac{2}{H_0}\right) e^{10\varphi/3} \left[D^i_{\ j} \varphi + 
\frac{5}{3} \left(\varphi^{,i} \varphi_{,j} - \frac{1}{3} 
\left(\nabla \varphi\right)^2
\delta^i_{\ j} \right) \right] . 
\end{eqnarray}

Let us first explain how the renormalization group method works
at the level of two gradients starting from the solution of Eq.\
(\ref{psi}) written in the form
\begin{equation}
\label{mu}
\Psi = \Psi_{\rm in} +  
 e^{2\Psi_{\rm in}}\epsilon\left(a-a_{\rm in}\right), \qquad
\epsilon\equiv \frac{1}{18}\left(\frac{1}{a_0}\right)
\left(\frac{2}{H_0}\right)^2\left[\nabla^2 \varphi - \frac{5}{6} 
\left(\nabla \varphi\right)^2\right]  , 
\end{equation}
where $\epsilon\ll 1$. By taking the limit $\epsilon \ll 1$ we can
isolate the long-wavelength part of $\Psi$, in other words,
$\Psi_\ell$.

The constant $\Psi_{\rm in}$ represents the value of the gravitational
potential at some initial instant of time when the scale factor is
$a_{\rm in}$. We regularize the ${\cal O}(\epsilon)$ secular term by
introducing an arbitrary ``scale factor'' $\mu$ and a renormalized
constant $\Psi_{\rm in}=\Psi_R(\mu)+\epsilon\,\delta \Psi(\mu,a_{\rm
in})$.  If we split the term $(a-a_{\rm in})$ into $(a-\mu +\mu
-a_{\rm in})$, then to first order in $\epsilon$
\begin{equation}
\Psi=\Psi_R(\mu)+\epsilon \delta \Psi(\mu,a_{\rm in}) +
\epsilon\,e^{2\Psi_R(\mu)}(a-\mu+\mu-a_{\rm in}) .
\end{equation}
The counterterm $\delta \Psi$ is determined in such a way to absorb
the $(\mu -a_{\rm in})$-dependent term in the gravitational potential
$\Psi$:
\begin{equation}
\delta \Psi(\mu,a_{\rm in}) + e^{2\Psi_R(\mu)}(\mu-a_{\rm in})=0 .
\end{equation}
This defines the renormalization-group transformation
\begin{equation}
\Psi_R(\mu)=\Psi_{\rm in} + \epsilon\, e^{2\Psi_R(\mu)}(\mu-a_{\rm in})  ,
\end{equation}
and the renormalization group equation
\begin{equation}
\label{rge}
\frac{\partial \Psi_R(\mu)}{\partial\mu}=\epsilon\,e^{2\Psi_R(\mu)}  .
\end{equation}
The solution of Eq.\ (\ref{rge}) is
\begin{equation}
\Psi_R(\mu)=-\frac{1}{2}\ln\left(c_2-2\,\epsilon\,\mu\right)  ,
\end{equation}
where $c_2=e^{-10\varphi/3}$ is the constant of integration.  
Equating $\mu$ to the generic scale factor we find that
the renormalized improved solution for the gravitational potential
at the level of two gradients is given by
\begin{equation}
\label{rgeimproved}
\Psi = \Psi_R\left(\mu=a\right)=\frac{5}{3}\varphi  -\frac{1}{2} \ln\left[1-
\frac{1}{9}\left(\frac{a}{a_0}\right)
\left(\frac{2}{H_0}\right)^2
 e^{10\varphi/3}
\left(\nabla^2 \varphi - \frac{5}{6} 
\left(\nabla \varphi\right)^2\right)\right]  . 
\end{equation}

If expanded up to two gradients, this solution coincides with Eq.\
(\ref{psi}).  Since by construction one should take the
long-wavelength part of the argument of the logarithm in the solution
of Eq.\ (\ref{rgeimproved}), it is easy to see that the latter matches
Eq.\ (\ref{psilong}) expanded up to two gradients.  Indeed, write Eq.\
(\ref{psilong}) as
\begin{equation}
\label{psilong2}
\Psi_\ell({\bf x},t)- \Psi_\ell({\bf x},t_{\rm in}) =- \frac{1}{3} \ln 
\left(\frac{\int_D \, e^{-3\Psi}\,d^3x}{\int_D \,
 e^{-3\Psi_{\rm in}} d^3x}
 \right)   .
\end{equation}
Inserting Eq.\ (\ref{psi}) into Eq.\ (\ref{psilong2}) and
expanding up to two gradients using the fact that
$\Psi_\ell({\bf x},t_{\rm in})=5\varphi/3$, we obtain
\begin{eqnarray} 
\Psi_\ell({\bf x},t)- \Psi_\ell({\bf x},t_{\rm in}) &\simeq&- \frac{1}{3} \ln 
\left(\frac{\int_D \, e^{-3\Psi_{\rm in}}\left(1-3\,e^{10\varphi/3}\,
\epsilon \,a\right)
\,d^3x}{\int_D \, e^{-3\Psi_{\rm in}}\,d^3x}\right)\nonumber\\
&\simeq & - \frac{1}{3} \ln \left(1-3
\frac{\int_D \, e^{-3\Psi_{\rm in}}\,e^{10\varphi/3}\,
\epsilon\, a\,d^3x}{\int_D \, e^{-3\Psi_{\rm in}}\,d^3x}
\right)\nonumber\\
&\simeq& -\frac{1}{2}\ln\left(1-\left\langle 
 \frac{1}{9}\left(\frac{a}{a_0}\right)
\left(\frac{2}{H_0}\right)^2
 e^{10\varphi/3}
\left[\nabla^2 \varphi - \frac{5}{6} 
\left(\nabla \varphi\right)^2\right] 
\right\rangle_{D_{\rm in}}
\right)  ,
\end{eqnarray}
which coincides with Eq.\ (\ref{rgeimproved}).  Notice, in particular,
that Eq.\ (\ref{rgeimproved}) differs from the toy gravitational
potential adopted by Hirata and Seljak \cite{seljak}.

Let us compute the corresponding deceleration parameter 
\begin{equation}
\label{de}
q=-\frac{\dot{H}_D}{H_D^2}-1=-\frac{\dot H-\ddot{\Psi}_\ell}{\left(
H-\dot{\Psi}_\ell\right)^2}  ,
\end{equation}
where we have used the fact that
 $H_D=H-\dot{\Psi}_\ell$. Inserting Eq.\ (\ref{rgeimproved})
into Eq.\ (\ref{de}), we find that at late times 
\begin{equation}
\label{kkkk}
q\sim\frac{3}{2}\cdot \frac{2}{3}-1=0,
\end{equation}
{\it i.e.,} the deceleration parameter tends to zero.  This result
confirms our expectation that at the (resummed) lowest order in the
gradient expansion, the Universe turns out to be curvature-dominated
at late times, which is equivalent to a Universe with effective
equation of state $w=-1/3$. What about the resummation of the 
long-wavelength perturbations at higher order in gradient terms?  The
curvature term is a series of gradients, and can be written in the
form
\begin{equation}
\label{k1}
R=\sum_{n\geq 1}\,e^{2n \Psi_{\rm in}}\,c_n\,a^{n-2}  ,
\end{equation}
where $c_n={\cal O}\left(\partial^{2n}\right)$ is a coefficient
containing $2n$ gradients.  Repeating the resummation procedure
outlined for the case of two gradients, one can easily show that at
any given order $n$ the renormalized solution reads
\begin{equation}
\Psi_R(a)\sim -\frac{1}{2n}\ln \left(1-2 n\,c_n\, a^n\right)  .
\end{equation}
Since at very late times $\Psi_R\sim -\frac{1}{2}\ln\,a$, the
corresponding deceleration parameter at late times goes like in Eq.\
(\ref{kkkk}). We conclude that at any order in the gradients, at late
times the effect of the resummation of the long-wavelength
perturbations is simply to generate a curvature term. This conclusion
may be obtained also by inspecting Eq.\ (\ref{k1}) after the constant
of integration $\Psi_{\rm in}$ has been promoted to the renormalized
quantity $\Psi_R$. Each term in the series gives a contribution to $R$
which scales as $a^{-3}\sim t^{-2}$.  If only long-wavelength
perturbations were present, the true scale factor $a_D$ would scale
like
\begin{equation}
a_D=a\,e^{\frac{1}{2}\ln\, a}=a^{3/2}  ,
\end{equation}
and $\langle R\rangle_D$ would scale like $a_D^{-2}$. Therefore, if
{\it only} long-wavelength perturbations were present, at large times 
and at any order in the gradients 
the line-element would take the form of a curvature dominated
Universe, with $h_{ij}\sim t^2\,C_{ij}({\bf x})$, where $C_{ij}({\bf
x})$ is a function of spatial coordinates only. 

In summary, super-Hubble perturbations cannot be distinguished 
from the background for local observers. Thus a Universe which is pure 
matter and has only super-Hubble perturbations, looks like a FRW universe 
to the local observer. 
Even if we started with a flat Universe plus perturbations, it is clear 
that the local observer will interpret what she/he sees as a FRW model with 
curvature (it would need a fine tuning to have k=0 within the Hubble patch).
Now, as there is only matter and curvature in that model, 
the curvature will eventually dominate at late times, as a (open) 
non-flat matter Universe is dominated by curvature at late times.

\subsection{The effect of sub-Hubble modes}

Dealing with the backreaction of sub-Hubble perturbations, and
therefore attacking the issue of the cross-talk between the sub-Hubble
modes and the homogeneous mode, is more difficult than dealing with the
super-Hubble modes because, as we shall see, the gradient expansion
displays an instability of the perturbative series.  

In the effective Friedmann description of the inhomogeneous Universe one
wishes to compute the typical value of the local observables averaged
over the comoving volume $D$. By that we mean the ensemble average of
such a volume average. The cosmological perturbations are treated as
variables that take random values over different realizations of
volumes $D$. In other words, we calculate the typical value of a
quantity for a region of given size as the statistical mean over many
different similar regions. This typical value is generically
accompanied by a variance. If the size of the comoving volume $D$ is
much smaller than the global inflationary volume, then we can imagine
placing this volume in random locations within a region whose size is
much bigger than the size of $D$. By the ergodic property, this is
equivalent to taking random samples of the ensemble for a fixed
location of the box. In other words, one can replace the expected
value of a given quantity averaged over a given comoving volume $D$
with the ensemble average of the volume average, denoted by
$\overline{\langle \cdots\rangle}_D$. Since we are interested in the
role of sub-Hubble perturbations which cause a tiny variation of the
value of the gravitational potential from one Hubble patch to another,
the variance of the local mean observables is small. Under these
circumstances, we can safely replace the spatial average with the
ensemble average. This automatically implies that the perturbations
which contribute to the effective dynamics are no longer restricted to
receive contributions peaked at  modes comparable to the
Hubble-size  (technically, this
means that the window filter function defining the size of the
comoving volume $D$ plays no role) and therefore can be much bigger
than of order $10^{-5}$ (or powers of it)\footnote{It is
important to stress that
although the evolution of the kinematical backreaction and the
mean curvature are  obtained for
    averaged fields restricted to the domain $D$, the solutions to the
    averaged equations are actually influenced by inhomogeneities
    outside the domain  $D$ too, since the initial data are to be constructed
    non-locally and so take the fields on the whole Cauchy hypersurface
    into account. We thank T. Buchert for discussions on this issue.}.

Let us start by considering the lowest order in a gradient expansion,
{\it i.e.,} keeping only two spatial derivatives.  The mean local
curvature will be non-vanishing, but $Q_D$ will be zero at this order,
as $\dot \Psi$ contains at least two spatial derivatives. In such a
case, the integrability relation Eq.\ (\ref{relation}) immediately
shows that the only consistent solution is
\begin{equation}
\label{rw}
\langle R\rangle_D\propto a_D^{-2}  ,
\end{equation}
{\it i.e.,} the effect of sub-Hubble perturbations at this order is to
generate a standard curvature-like term in the effective Friedmann
equations, scaling as the inverse square of the scale factor. This
simple result holds at any order in perturbation theory (provided that
one keeps only two spatial derivatives) and represents a
straightforward extension of what found   in Ref.\ \cite{chung} where
it was shown  that to 
second order in spatial gradients and in the gravitational
potential, cosmological perturbations amount only to a renormalization
of the local spatial curvature (this result valid up to two gradients
was though improperly  applied to the findings of Ref.\
\cite{KMNR}, where more than two spatial derivatives were included, for
instance through the physical redshift; this point was also noticed
in Ref.\ \cite{rasanen}).  The result of Eq.\ (\ref{rw}) is reminiscent
of the so-called {\it vacuole} model (see, {\it e.g.,} Ref.\
\cite{hammer}).\footnote{We thank S. Carroll for correspondence on
this issue.}  Consider indeed a spherical region of a perfectly
uniform Universe.  Suppose that the matter inside that spherical
region is squeezed into a smaller uniform spherical distribution with
higher density. By mass conservation there will be a region in
between the overdense sphere and the external Universe that is
completely empty. Einstein's equation are exactly solvable for this
situation in terms of the Tolman-Bondi metric.  The outside FRW
Universe is totally unaffected by such a rearrangement.  By Birkhoff's
theorem, the empty shell will be described by the Schwarzschild
metric. Finally, the interior region will behave like a homogeneous
and isotropic FRW Universe, but with different values of the
cosmological parameters. These parameters will exactly obey the
conventional Friedmann equations, and someone who lived inside there
would have no way of telling that those parameters did not describe
the entire Universe.

However, the situation changes if we consistently account for
higher-order derivative terms both in $\langle R\rangle_D$ and in
$Q_D$. The lowest non-zero contribution to $Q_D$ contains four spatial
gradients and goes like $a^2H^2 \propto a^{-1}$; this corresponds to a
similar term with four gradients in the mean spatial curvature.

Let us further elaborate on these findings.  In all generality one can
write
\begin{eqnarray}
Q_D & = & \sum_{n=2}^\infty q_n a^{n-3} \nonumber \\
\langle R \rangle_D & = &
\sum_{n=1}^\infty r_n a^{n-3}, 
\end{eqnarray}
where $q_n$ and $r_n$ are expansion coefficients containing $2n$
spatial gradients. (Note that $q_1=0$, as $Q_D$ starts from 4
gradients.)  One may wonder about the actual range of validity of the
gradient-expansion technique. At first sight it might appear to be
valid only to describe inhomogeneities on super-Hubble scales, {\it
i.e.,} for comoving wave-numbers $k \lesssim aH$.  However, this is
not really the case!  As one can easily check, terms of order $n$ in
the expansion, {\it i.e.,} terms with $2n$ gradients, contain the
peculiar gravitational potential $\varphi$ to power $m$ with $2n\geq m
\geq n$. The dominant contribution at each order $n$ ({\it i.e.,} with
$2n$ gradients) is Newtonian, {\it i.e.,} coming from terms of the
type $(\partial^2 \varphi)^n$. However, these terms both in $Q_D$ and
$\langle R\rangle_D$ sum up to produce negligible surface terms when
averaged over a large volume, so that the leading terms become the
first post-Newtonian ones, {\it i.e.,} those proportional to
$(\partial^2 \varphi)^{n-1} (\partial \varphi)^2$. In other words the
expansion is shielded from the effect of the Newtonian terms, which
could in principle be almost arbitrarily large, by the volume
averaging. The same {\it protection} mechanism, however, does not
apply to the non-Newtonian terms in the expansion, simply because they
cannot be recast as surface terms.  This simple reasoning immediately
leads to the conclusion that the actual limit of validity of our
expansion at order $n$, is set by $(k/aH)^{2n} \varphi^{n+1} \lesssim
1$. Because of the nearly-Gaussian nature of our inflationary seed
$\varphi$, it is clear that the lowest-order term able to produce a
big contribution to $Q_D$ and $\langle R \rangle_D$ appears for $n=3$,
{\it i.e.,} a term with six gradients.  The importance of the
six-derivative post-Newtonian terms has indeed been stressed also by
Notari \cite{alessio}.  It is a disconnected fourth-order moment of
$\varphi$ of the type
\begin{equation}
\left\langle(\nabla^2 \varphi)^2/H_0^4\rangle \langle (\nabla
\varphi)^2/H_0^2 \right\rangle ,
\label{six}
\end{equation}
having assumed that the spatial average coincides with the ensemble
average. At this level an {\it instability} of the perturbative
expansion is produced by the combination of the small post-Newtonian
term $\langle (\nabla \varphi)^2/H_0^2 \rangle$ (of order $10^{-5}$)
with the Newtonian term $\langle(\nabla^2 \varphi)^2/H_0^4\rangle$, which can
be almost arbitrarily large \cite{oldKMNR}, due to the logarithmic
dependence on the ultraviolet cut-off (for a scale-invariant spectrum
and cold dark matter transfer function).\footnote{Unlike the standard
perturbative approach, we need not require small matter density
fluctuations $\nabla^2 \varphi/H_0^2 \ll 1$, in our approach.  This is
because in the evaluation of mean observables, powers of $\nabla^2
\varphi/H_0^2$ either give rise to tiny surface terms or get
multiplied by the small post-Newtonian term
$(\nabla\varphi)^2/H_0^2$.}

It is important to stress that the six derivative terms give a
contribution to $Q_D$ which scales like $a^3H^2=$ constant; similarly
the six-gradients contribution to the smoothed curvature scales like
$a^2/a^2=$ constant. So these terms give rise to a sizeable {\it
effective cosmological constant-like term} in our local Friedmann
equations.  In order to estimate correctly the six-gradients terms one
needs the metric coefficients $\Psi$ and $\chi^i_{\ j}$ up to four
gradients (whose explicit expressions are given in the Appendix). 

The existence of a large contribution at six gradients, however,
suggests that higher-order gradient terms will similarly lead to
large corrections to the FRW background expansion rate. This is indeed
the case.  In the large-$n$ limit there will be large contributions
coming from perturbations in the quasi-linear regime ($|\delta_{FRW}|
\gtrsim 1$). These generic conclusions, however, also tell us that
stopping the expansion at six gradients would be completely arbitrary
and that, in any case, the perturbative approach cries for a more
refined treatment than simply counting powers of the scale factor as
done in Ref.\ \cite{alessio}.  The existence of large corrections to
the background should be taken strictly as evidence for an instability
of the FRW background caused by nonlinear structure formation in the
Universe.  The actual quantitative evaluation of their effect on the
expansion rate of the Universe would however require a truly
non-perturbative approach, which is clearly beyond the aim of this
paper.

Connected to this fact is a technical obstacle in extending the
validity of the gradient expansion to late times and/or to the
nonlinear regime. This comes from the fact that the metric determinant
may become negative, indicating an internal inconsistency of the
approximation. In Ref.\ \cite{4grad} the problem is solved by using an
``improved'' approximation scheme which expresses the metric as a
``square;'' this choice guarantees non-negativity and leads to a GR
extension of the classical Zel'dovich approximation of Newtonian
theory. It is then shown that with suitable choice of the initial
seed, such an improved approximation provides an excellent match to
an exact inhomogeneous solution of Einstein's field equations, the so
called {\it Szekeres metric} \cite{szek}, which describes locally
axisymmetric (pancake) collapse of irrotational dust. Alternatively,
exploiting the non-perturbative continuity equation,
$(1+\delta_{\rm FRW})^{-1}=\int_{t_{\rm in}}^t\, dt\, \theta$, 
one can easily convince her/himself that the
determinant of the metric is always well-defined.

In order to take one step forward in the gradient-expansion approach,
we will use the same renormalization group technique previously
applied to deal with the backreaction of super-Hubble modes.  Let us
start by dealing with the case of two gradients.  Can we apply the
renormalization technique to the case of sub-Hubble perturbations up
to two gradients? The answer is yes, since the spatial averages of
objects like $\nabla^2\varphi/H_0^2$ and
$\left(\nabla\varphi\right)^2/H_0^2$ can be replaced by the
corresponding ensemble averages and therefore are tiny (of the order
of $10^{-5}$).  The renormalized growing solutions at two
gradients reads therefore\footnote{The long-wavelength part of the
factor $e^{5\varphi({\bf x})/3}$, associated with each spatial
gradient, can be re-absorbed by a redefinition of the spatial
coordinates, as noticed in Ref.\ \cite{seljak}, and does not play any
role when evaluating ensemble averages as well as in the backreaction
problem (we thank M.\ Porrati for correspondence on this issue); the
small-wavelength part, on the other hand, can be expanded as
$(1+5\varphi_s/3)\sim 1$ because $\varphi_s \sim 10^{-5}$.
Nonetheless, we prefer to show them explicitly because they provide
the initial condition $C_{\rm in}$ for the renormalization approach.}
\begin{eqnarray}
\label{subrge}
\Psi & = &\frac{5}{3}\varphi  -\frac{1}{2} \ln\left[1-
\frac{1}{9}\left(\frac{a}{a_0}\right)
\left(\frac{2}{H_0}\right)^2
 e^{10\varphi/3}
\left(\nabla^2 \varphi - \frac{5}{6} 
\left(\nabla \varphi\right)^2\right)\right] ,
\nonumber\\
\chi^i_{\ j} & = & -\frac{1}{2} \ln\left\{1-
 \frac{2}{3}\left(\frac{a}{a_0}\right)
\left(\frac{2}{H_0}\right)^2 e^{10\varphi/3}
\left[D^i_{\ j} \varphi + \frac{5}{3} 
\left(\varphi^{,i} \varphi_{,j} - \frac{1}{3} \left(\nabla \varphi\right)^2
\delta^i_{\ j} \right) \right] \right\} .
\end{eqnarray}
The next step consists in solving for the cosmological perturbations
at four gradients. The equations of motion for the gravitational
potential $\Psi$ and for $\chi^i_{\ j}$ at four gradients are given by
Eqs. (\ref{k}) and (\ref{kk}) where the sources in the right-hand-side
are computed inserting the solutions of Eq.\ (\ref{subrge}).

Upon defining the coefficient
\begin{equation}
{\cal E} =  \frac{1}{9}e^{10\varphi/3}
\left(\nabla^2 \varphi - \frac{5}{6} 
\left(\nabla \varphi\right)^2\right)  ,
\end{equation}
the matrix
\begin{equation}
{\cal F}^i_{\ j}=\left(
\partial^i\partial_j\varphi-\frac{5}{6}\partial^i\varphi\partial_j\varphi
\right)  ,
\end{equation}
and the traceless matrix
\begin{equation}
{\cal E}^i_{\ j}=\frac{2}{3}e^{10\varphi/3}
\left[D^i_{\ j} \varphi + \frac{5}{3} 
\left(\varphi^{,i} \varphi_{,j} - \frac{1}{3} \left(\nabla \varphi\right)^2
\delta^i_{\ j} \right) \right]  ,
\end{equation}
the growing solution at four gradients for the gravitational potential
assumes the form 
\begin{eqnarray}
\label{ss}
\Psi &\simeq& -
\frac{1}{24}{\rm Tr}\,\left[\ln \left(1-\left(\frac{a}{a_0}\right)
\left(\frac{2}{H_0}\right)^2
{\cal E}\right)\, \ln\left(1-\left(\frac{a}{a_0}\right)
\left(\frac{2}{H_0}\right)^2
{\cal E}\right)\right]
\nonumber\\
&+&\frac{5}{36\,{\cal E}}\,
{\rm Tr}\left[\ln\left(1-\left(\frac{a}{a_0}\right)
\left(\frac{2}{H_0}\right)^2
{\cal E}\right)\ln\left(1-\left(\frac{a}{a_0}\right)
\left(\frac{2}{H_0}\right)^2
{\cal E}\right)\,{\cal F}\right]  .
\end{eqnarray}
A similar solution can be obtained for $\chi^i_{\ j}$ at four
gradients. At late times the solution for the gravitational potential
grows like $(\ln a)^2$. A renormalization procedure can be applied to
the solutions at four gradients because there are still ``small''
perturbative terms at hand, for instance terms like
$\left(\nabla\varphi\right)^4$ whose spatial average is small. This
amounts to saying that one has to take the solution of Eq.\
(\ref{ss}), expand the arguments of the logarithms and apply the
renormalization procedure described previously at second order in the
``perturbative parameter'' ${\cal E}$.  The renormalized solution for
the gravitational potential will grow like $(\ln a)$ at large times.

The lesson to learn from this computation is that, if we proceed
further and go to six gradients, the unrenormalized gravitational
potential, as well as $\chi^i_{\ j}$, will grow like $(\ln a)^3$. This is
surely a step forward compared to the simple counting of powers of the
scale factor which predicts that, at six gradients, the gravitational
potential should grow like $a^3$. However, at this stage the
renormalization procedure fails because it involves terms like the one
in Eq.\ (\ref{six}), which may be easily of order unity. Even the
resummed perturbative expansion shows an instability produced by the
combination of post-Newtonian and Newtonian terms; solutions with $2n$
gradients are expected to behave like $(\ln a)^n$.  If taken at face
value, such a time-behavior of the gravitational potential would lead
to acceleration of our local Hubble patch. To put this indication on
firmer grounds (or to disprove it), however, one should go
beyond the perturbative approach adopted in this paper. Our result may
spur the efforts toward the search for a nonperturbative description
of the dynamics of the system which would account for combinations of
large Newtonian and small post-Newtonian terms.

As a concluding remark of this subsection, we address a  
common objection to the use of the synchronous and 
comoving gauge in addressing the backreaction problem, namely that 
the occurrence of shell-crossing singularities (caustics) 
in the evolution of collisionless fluids might prevent 
the analysis to be carried over into the fully non-linear regime. 
We would like to point out that the instability we find
in the gradient expansion is unrelated to 
shell-crossing singularities. 
This can be immediately appreciated by noting that: i) shell-crossing 
instabilities imply the emergence of {\it divergent} gradients terms, 
while our instability shows up through an infinite number of {\it finite} 
gradient terms; ii) shell crossing is well known to lead to an infinite 
Newtonian term, while our effect involves a tiny Newtonian term. 
It should also be stressed that the occurrence of caustics 
does not represent a serious limitation of our approach; indeed, 
the very fact that caustics only carry a small amount of mass implies 
that they can be easily smeared over a finite region out in such a way that 
their presence does not affect the mean expansion rate of the Universe.
 
For the sake of completeness, in the next subsection we will 
address the problem at hand within the commonly used weak-field 
approximation in the Poisson gauge.

\subsection{The backreaction in the weak-field approximation}

So far, in evaluating the effect of backreaction we have been making
use of a perturbative approach in which non-linear dynamical
quantities are explicitly expressed in terms of the inflationary seed
perturbation $\varphi$ and its spatial derivatives. This is the reason
why higher and higher gradients of $\varphi$ appear in our
results. The same conclusion would hold also in different gauges as
well as by using different perturbative schemes.

One might however wonder whether the back-reaction problem can be
approached {\it directly} in terms of non-linearly evolved
variables. Related to this issue is the gauge choice.
Non-perturbative approaches are indeed possible both in the comoving
gauge adopted so far (see Ref.\ \cite{mater}), and in the more
commonly used Poisson gauge \cite{bert}.  Working out the effects of
back-reaction in a non-comoving gauge is indeed fully legitimate,
provided a well-defined space-time splitting is performed, {\it e.g.,}
by means of the ADM approach \cite{buchert2}.  We will here only
sketch how back-reaction effects can be evaluated in the Poisson
gauge, leaving to a subsequent paper a more detailed and quantitative
analysis of the problem.
  
The line-element of the Poisson gauge reads (see, {\it e.g.,} Ref.\
\cite{carbone})
\begin{equation}
\label{poissonmetric}
ds^2 = a^2(\tau)\left\{ - \left(1+ 2 \phi_P \right) 
d\tau^2 - 2 V_i d\tau dx^i 
+ \left[\left(1-2 \Psi_P\right)\delta_{ij} + h^{(T)}_{ij} \right] dx^i dx^j 
\right\} . 
\end{equation}
where $\tau$ is the conformal time and $a(\tau) \propto \tau^2$ is the
FRW background scale-factor for our irrotational dust source.  It is
important to stress that this line-element is meant to include
perturbative terms of any order around the FRW background. The
quantities $V_i$ are pure vectors, {\it i.e.,} they are divergenceless,
$\partial^i V_i=0$, while $h^{(T)}_{ij}$ represent traceless and
transverse ({\it i.e.,} pure tensor) modes, $h^{(T)i}_i = \partial^i
h^{(T)}_{ij}=0$ (spatial indices are raised by the Kronecker
symbol). Vector and tensor metric modes are, respectively, of ${\cal
O}(1/c^3)$ and ${\cal O}(1/c^4)$. To leading order in powers of $1/c$
the above line-element is known to take the well-known {\it
weak-field} form
\begin{equation}
\label{weakfield}
ds^2 = a^2(\tau)\left[ - (1+ 2 \phi_P) d\tau^2 + 
\left(1-2 \psi_P\right)\delta_{ij} d x^i d x^j \right] ,
\end{equation}
where the scalars $\phi_P$ and $\psi_P$ are both ${\cal O}(1/c^2)$ and
$\phi_P=\psi_P=\Phi_N/c^2$; the Newtonian gravitational
potential $\Phi_N$ is related to density fluctuations $\delta\rho$ by
the cosmological Poisson equation $\nabla^2 \Phi_N = 4\pi G a^2
\delta\rho$.  It is easy to realize that this form is accurate enough
to describe structure formation within the Hubble radius as long as
the considered wavelengths are much larger than the Schwarzschild
radius of collapsing bodies \cite{peebles}. 

The crucial point is that the kinematical back-reaction will contain
the relevant term \cite{buchert2} $\langle N^2 \Theta^2 \rangle_D$,
where $\Theta = u^\mu_{\ ;\mu}$ ($u^\mu$ being the fluid
four-velocity). Here $N$ is the inhomogeneous lapse function needed to
express the Poisson-gauge coordinate time $t_P = \int d\tau \;a(\tau)$
as a function of the proper time $t$ of comoving observers. This issue
was already pointed out in Ref.\ \cite{oldrasanen} where an approximate
explicit expression for $N$ was given (a second-order perturbative
expression can be found in Ref.\ \cite{oldKMNR}); a term of the type
$\left(\nabla \Phi_N \right)^2$ appears explicitly in $N$.  What is
important for us here is that $Q_D$ will clearly display the same type
of post-Newtonian (hence non-total derivative) terms which were found
in the comoving gauge using our gradient expansion, namely terms of
the type
\begin{equation}
\label{p}
\left\langle \left(\nabla^2 \Phi_v\right)^2 \left(\nabla\Phi_N \right)^2 
\right\rangle , 
\end{equation}
where $\Phi_v$ is the velocity potential, which coincides (up to a
sign) with the gravitational potential $\Phi_N$ on linear scales; more
generality $\Phi_v$ and $\Phi_N$ are connected by a cosmological
Bernoulli equation (see, {\it e.g.,} Ref.\ \cite{mater}).  Similar
terms appear in the mean curvature when projecting onto the comoving
observer frame.  We stress again that the terms of the type
(\ref{p}) appear only when considering the correct effective
description of the average dynamics which has to include
the kinematical backreaction term. Notice that this does not
amount to saying that post-Newtonian effects are relevant in the
dynamical evolution of the gravitational and velocity potentials themselves. 
Indeed the  expression (\ref{p}) requires evaluation of the generally
non-linear potentials $\Phi_v$ and $\Phi_N$ which may be readily
obtained through the use of standard $N$-body simulations. 
Owing to the non-linear (hence non-Gaussian) nature
of the   potentials $\Phi_v$ and $\Phi_N$, the  average (\ref{p}) contains
both a disconnected term, as in our previous treatment, and a non-zero
connected four-point moment which is dominated by mildly non-linear
scales, of order a few Mpc.  

Contrary to what happens in the
synchronous gauge when the result is expressed in terms of the initial
seed $\varphi$, in the weak-field approximation the number of
gradients is expected to be finite and the complexity of the problem resides 
in the non-perturbative evaluation of the evolved potentials 
$\Phi_v$ and $\Phi_N$. 

It is interesting to note that the combination in Eq.\ (\ref{p}) provides a
contribution to $Q_D$ which is of the order of $H^2$ and, using the
linear dependence, nearly constant in time.

\section{Conclusions \label{conclusions}}

The most astonishing recent observational result in cosmology is the
indication that our Universe is presently undergoing a phase of
accelerated expansion. One possible explanation of the observations is
that the Universe is homogeneously filled with a fluid with negative
pressure that counteracts the attractive gravitational force of matter
fields. Another possible explanation is a modification of GR on large
distance scales.

In this paper we have elaborated on the alternative idea that the backreaction
of cosmological perturbations may cause the cosmic acceleration 
\cite{KMNR,oldKMNR,japan,oldrasanen,rasanen,alessio,sw,bmr,rtolman}.  Following
Buchert \cite{buchert1,buchert2}, we have provided the effective
Friedmann equations describing an inhomogeneous Universe after
smoothing. The effective dynamics is governed by two scalars: the
so-called kinematical backreaction $Q_D$ and the mean spatial
curvature $\langle R\rangle_D$. They enter in the expression for the
effective energy density and pressure in the Friedmann equations
governing the mean evolution of a local domain $D$.  For positive
$Q_D$, acceleration in our local Hubble patch may be attained despite
the fact that fluid elements cannot individually undergo accelerated
expansion.  Indeed, the very fact that the smoothing process does not
commute with the time evolution invalidates the no-go theorem, which
states that there can be no acceleration in our local Hubble patch if
the Universe only contains irrotational dust.

Through the renormalization group technique, we have then shown that
super-Hubble modes can be resummed at any order in perturbation theory
yielding a  local curvature term $\sim a_D^{-2}$ at large times.  
We then turned our attention to
the backreaction originating from modes within our Hubble radius,
studying perturbatively their time-behavior. In this case our findings
indicate that an instability occurs in the perturbative expansion,
which may be not taken care of by the renormalization group procedure
since terms of the form $H^2\langle\delta_{\rm
FRW}^2(v/c)^2\rangle$ (where $v$ is the peculiar velocity)
start appearing both in $Q_D$ and in the mean spatial curvature. Such
terms are not as small as order $10^{-5}\,H^2$; on the contrary the
averaging procedure allows the combination of post-Newtonian and
Newtonian terms to acquire values of order $H^2$.  Since the
perturbation approach breaks down, we may not predict on firm grounds
that backreaction is responsible for the present-day acceleration of
the Universe.  However, it is intriguing that such an instability
shows up only recently in the evolution of the Universe and that this
picture is further supported by a very general result; 
as shown explicitely by Buchert et al. \cite{bks}, even a tiny
back-reaction term can drive the cosmological parameters on the
averaging domain far away from their global values of the standard FRW
model, thus modifying the global expansion history of the Universe. 
Other aspects of the scenario discussed in this
paper, such as the dynamics of perturbations on observable scales,
will be the subject of a forthcoming publication.

\appendix*
\section{Fourth-order gradient-expansion approximation to the solution of
Einstein's field equations}

For completeness, we give here the explicit expression for $\Psi$ and
$\chi^i_{\ j}$ up to four gradients (we refer the reader to Ref.\
\cite{pilli} for the detailed derivation of these results).  We have
\begin{eqnarray}
\label{psi4}
\Psi & = & \frac{5}{3}\varphi + \frac{1}{18}\left(\frac{a}{a_0}\right)
\left(\frac{2}{H_0}\right)^2
 e^{10\varphi/3}\left[\nabla^2 \varphi - \frac{5}{6} 
\left(\nabla \varphi\right)^2\right] + 
\frac{1}{504}\left(\frac{a}{a_0}\right)^2
\left(\frac{2}{H_0}\right)^4  
e^{20\varphi/3} \nonumber \\
& & \times \left[\frac{23}{9} 
\left(\nabla^2 \varphi\right)^2 - \frac{10}{3}
\varphi^{,ij}\varphi_{,ij} - 
\frac{100}{9} \varphi_{,i} \varphi_{,j} \varphi^{,ij}
+ \frac{35}{27} \nabla^2 \varphi \left(\nabla \varphi\right)^2 -
\frac{1675}{324} 
\left(\nabla \varphi\right)^2 \left(\nabla \varphi\right)^2
\right] \\
\label{chi4}
\chi^i_{\ j} & = & - \frac{1}{3}\left(\frac{a}{a_0}\right)
\left(\frac{2}{H_0}\right)^2 e^{10\varphi/3}
\left[D^i_{\ j} \varphi + \frac{5}{3} 
\left(\varphi^{,i} \varphi_{,j} - \frac{1}{3} \left(\nabla \varphi\right)^2
\delta^i_{\ j} \right) \right] \nonumber \\
&  & + \frac{1}{504}\left(\frac{a}{a_0}\right)^2
\left(\frac{2}{H_0}\right)^4 e^{20\varphi/3} 
\left\{38  \left( {\varphi}^{,ki}{\varphi}_{,kj} 
-\frac{1}{3} {\varphi}_{,kl} {\varphi}^{,kl}  \delta^{i}_{\ j}\right) \right.
\nonumber \\
& & - \frac{128}{3} \left[(\nabla^{2}\varphi)  \varphi^{,i}_{\ ,j} 
- \frac{1}{3} (\nabla^{2}\varphi)^{2} \delta^{i}_{\ j}\right] 
+ \frac{890}{27} (\nabla^{2}\varphi) (\nabla\varphi)^{2} \delta^{i}_{\ j} 
- \frac{250}{9} (\nabla\varphi)^{2}  \varphi^{,i}_{\ ,j}  \nonumber \\
& & - \frac{640}{9}(\nabla^{2}\varphi) \varphi^{,i} \varphi_{,j}
-\frac{380}{9}\varphi_{,k}\varphi_{,l}\varphi^{,kl}\delta^{i}_{\ j}
+ \frac{190}{3}\left( \varphi^{,ki}\varphi_{,k} \varphi_{,j} 
+ \varphi^{,i} \varphi_{,kj} \varphi^{,k} \right) \nonumber \\
& & \left. 
+ \frac{1600}{27} (\nabla\varphi)^{2} \left({\varphi}^{,i} \varphi_{,j}
- \frac{1}{3}(\nabla\varphi)^{2} \delta^{i}_{\ j}\right) \right\}  .
\end{eqnarray}

One can verify that these solutions  satisfy the energy constraint and the
momentum constraint up to the  relevant number of gradients.  It is also
important to stress that these expressions reproduce the  perturbative
second-order metric (see, {\it e.g.,} Ref.\ \cite{oldKMNR})  when only terms up
to second order in $\varphi$ are kept.

\begin{acknowledgments}
It is a pleasure to thank M.\ Bruni, T.\ Buchert, S.\ Carroll, D.\
Chung, D.\ Eisenstein, G.\ Ellis, G.\ Gelmini, A.\ Guth, J.\
Maldacena, Y.\ Nambu, P.\ Naselsky, A.\ Notari, J.\ Peebles, 
A.\ Pillepich, L.\ Pilo, M.\ Porrati, S.\ R\"as\"{a}nen, V.\ Sahni, 
R.\ Scherrer, U.\ Seljak, N.\ Straumann, G.\ Veneziano and 
L.\ Verde, for discussions during the various stages of this work.
\end{acknowledgments} 




\begin{thebibliography}{99}
\bibitem{comment}  
Bounds on the mean Hubble rate
and deceleration parameter from back-reaction have been
proposed in v3 of Ref.\ \cite{rasanen}. However, they apply only to those
regions which never turned around from the universal expansion, that is to 
those fluid elements for which the volume expansion rate never changed sign 
during the evolution.
\end{thebibliography}
\end{document}